 \documentclass[12pt,preprint]{aastex}

 \usepackage{emulateapj5}

\slugcomment{to appear in the Astrophysical Journal}

\shorttitle{CI Aquilae and RX J0513.9-6951}
\shortauthors{Hachisu and Kato}

\begin{document}

\title{A NEW CLUE TO THE TRANSITION MECHANISM BETWEEN OPTICAL HIGH AND
LOW STATES OF THE SUPERSOFT X-RAY SOURCE RX~J0513.9$-$6951, IMPLIED
FROM THE RECURRENT NOVA CI~AQUILAE 2000 OUTBURST MODEL}


\author{Izumi Hachisu}
\affil{Department of Earth Science and Astronomy, 
College of Arts and Sciences, University of Tokyo,
Komaba, Meguro-ku, Tokyo 153-8902, Japan} 
\email{hachisu@chianti.c.u-tokyo.ac.jp}

\and

\author{Mariko Kato}
\affil{Department of Astronomy, Keio University, 
Hiyoshi, Kouhoku-ku, Yokohama 223-8521, Japan} 
\email{mariko@educ.cc.keio.ac.jp}




\begin{abstract}
     We have found a new clue to the transition mechanism between optical 
high/X-ray off and optical low/X-ray on states of the LMC supersoft X-ray
source RX~J0513.9$-$6951.  A sharp $\sim 1$ mag drop is common to the
CI Aql 2000 outburst.  These drops are naturally attributed to 
cessation of optically thick winds on white dwarfs.
A detailed light-curve analysis of CI Aql indicates that the size 
of a disk drastically shrinks when the wind stops.
This causes $\sim 1-2$ mag drop in the optical light curve.  
In RX J0513.9$-$6951, the same mechanism reproduces sharp 
$\sim 1$ mag drop from optical high to low states.
We predict this mechanism also works on the transition from low 
to high states.  
Interaction between the wind and the companion star attenuates
the mass transfer and drives full cycles of low and high states.
\end{abstract}




\section{INTRODUCTION}
     The Large Magellanic Cloud (LMC) supersoft X-ray source 
RX~J0513.9$-$6951 (hereafter RX~J0513) shows a prominent
recurrency of optical high and low states with quasi-regular
intervals ($\sim 120$ days high and $\sim 40$ days low states).
Copious supersoft X-rays were detected only in the optical low states
\citep[e.g.,][]{sch93}.  The transition mechanism between
high and low states have not been fully elucidated yet, although a few
ideas were proposed \citep[see, e.g.,][for recent progress]{rei00}. 
We have found a new clue to this transition mechanism from modeling 
of the recurrent nova CI Aquilae.
\par
     CI Aql erupted in 2000 April has been densely observed in various
optical bands \citep*[e.g.,][]{hac02, kis01, led02, mat01, sch00} and
in X-ray bands \citep*{gre02}.  It reached the optical maximum 
($V \sim 9$ mag) on 2000 May 5.  The visual brightness
quickly decreased to $V \sim 13.5$ in about 50 days.  Then, it stays
at $V \sim 14$ for about 150 days, i.e., a plateau phase.
\par
     \citet{mat01} reported a sharp $\sim 2$ mag drop 
of $R_c$-magnitude on 2000 November 23, about 200 days after 
the optical maximum, and Kiyota (2001, VSNET archives, 
http://vsnet.kusastro.kyoto-u.ac.jp/vsnet/)
also observed a $\sim 2$ mag drop of $I_c$-magnitude
around the same day as Matsumoto et al.'s (2001) observation
\citep[see, e.g.,][]{hac02}.
Shortly after this drop, both the $R_c$- and $I_c$-magnitudes
recovered by $\sim 1$ mag but stayed at $\sim 1$ mag
below the level before the drop.  \citet{hac01ka} and \citet{hac02}
attributed this drop to cessation of massive winds by calculating 
time-evolution of optically thick winds on a $1.2~M_\sun$ 
white dwarf (WD).
\par
     The above feature of CI Aql light curve reminds us the optical 
high-to-low transition of RX~J0513.  In this paper, we try 
to reproduce the sharp drop in the optical light curves, 
both for CI Aql and RX~J0513, by the same mechanism.  
In \S 2, our numerical method and models for the CI Aql 2000
outburst are briefly introduced and summarized.  The numerical
results for the CI Aql 2000 outburst are given in \S 3. 
We try to reproduce the transition between high and low states of 
RX~J0513 in \S4.  Discussion follows in \S5.

\section{BASIC MODEL OF CI AQUILAE 2000 OUTBURST}
     The mid plateau phase is a common feature among the U Sco subclass
of the recurrent nova class, to which both U Sco and CI Aql belong.
The plateau phase is well explained by the contribution of an
irradiated disk \citep[e.g.,][]{hkkm00}.  The presence of the disk
in outburst was observationally confirmed by \citet{tho01}.
\par
     Our binary model is illustrated in Figure \ref{ci_aql_3figure},
which consists of a main-sequence star (MS) filling
its Roche lobe, a WD photosphere, and a disk around the WD.  
A circular orbit is assumed.
We also assume that the surfaces of the WD, the companion, and the disk 
emit photons as a blackbody at a local temperature 
which varies with position.
We assume an axi-symmetric accretion disk with the size and thickness of
\begin{equation}
R_{\rm disk} = \alpha R_1^*,
\label{accretion-disk-size}
\end{equation}
and
\begin{equation}
h = \beta R_{\rm disk} \left({{\varpi} 
\over {R_{\rm disk}}} \right)^\nu,
\label{flaring-up-disk}
\end{equation}
where $R_{\rm disk}$ is the outer edge of the optically thick 
part of the accretion disk,
$R_1^*$ is the effective radius of the inner critical Roche lobe 
for the WD component,
$h$ is the height of the surface from the equatorial plane, and
$\varpi$ is the distance from the symmetry axis.
Here, we adopt $\nu=1$ during the strong wind phase, but  
$\nu=2$ when rapid mass accretion resumes after the optically 
thick wind stops, to mimic the flaring-up disk 
by spray \citep*[e.g.,][]{sch97}.  
The other two parameters of $\alpha$ and 
$\beta$ are determined by light curve fittings.

\placefigure{ci_aql_3figure}

     We have used the same parameters as those in \citet{hac02}
unless otherwise specified.  The mass of the white dwarf was 
determined to be $1.2 \pm 0.05 ~M_\sun$ by \citet{hac01ka}.
The mass and temperature of the companion
star were estimated to be $\sim 1.5 M_\sun$ and 7,300 K 
\citep[see, e.g.,][for the present evolutional status]{hac01kb} 
from the light curve fitting of the orbital modulations in 
the 2000 outburst \citep{hac02, led02}.  
The inclination angle is a free
parameter, which is redetermined from our light curve fitting. 
\par
     Time evolution of the photospheric radius and temperature 
of the WD are calculated from Kato \& Hachisu's (1994) 
optically thick wind solutions \citep[see][for more details 
of the numerical methods]{hac01kb}.   When the photospheric radius
shrinks smaller than the binary size, the irradiation effects 
both of the disk and of the companion become important
\citep[see, e.g., Fig. 2 of][for such an example]{hkkm00}.

\section{WHAT WE LEARNED FROM CI AQUILAE}
     Orbital modulations during the plateau phase 
have been reported by \citet{mat01}.  Their $R_c$ light curve
data show a scatter of $\sim 0.1 - 0.2$ mag so that we average them 
into 36 phase bins as shown in Figure \ref{rc_mag_m15_wind_orbital}.
No secondary eclipses were observed in contrast to the later phase
orbital light curves \citep{led02}.  
To fit these orbital modulations as well as the 
entire evolution of $R_c$-magnitude, we have calculated light curves
with various parameters ($\alpha$ is changed by 0.1 steps, 
$\beta$ by 0.05 steps, and the inclination angle $i$ by 1$\arcdeg$ steps).
The best fit model is shown in Figure \ref{rc_mag_m15_wind_orbital},
i.e., $(\alpha, \beta)= (2.0, 0.05)$ and $i= 69\arcdeg$.

\placefigure{rc_mag_m15_wind_orbital}

\begin{figure}
\plotone{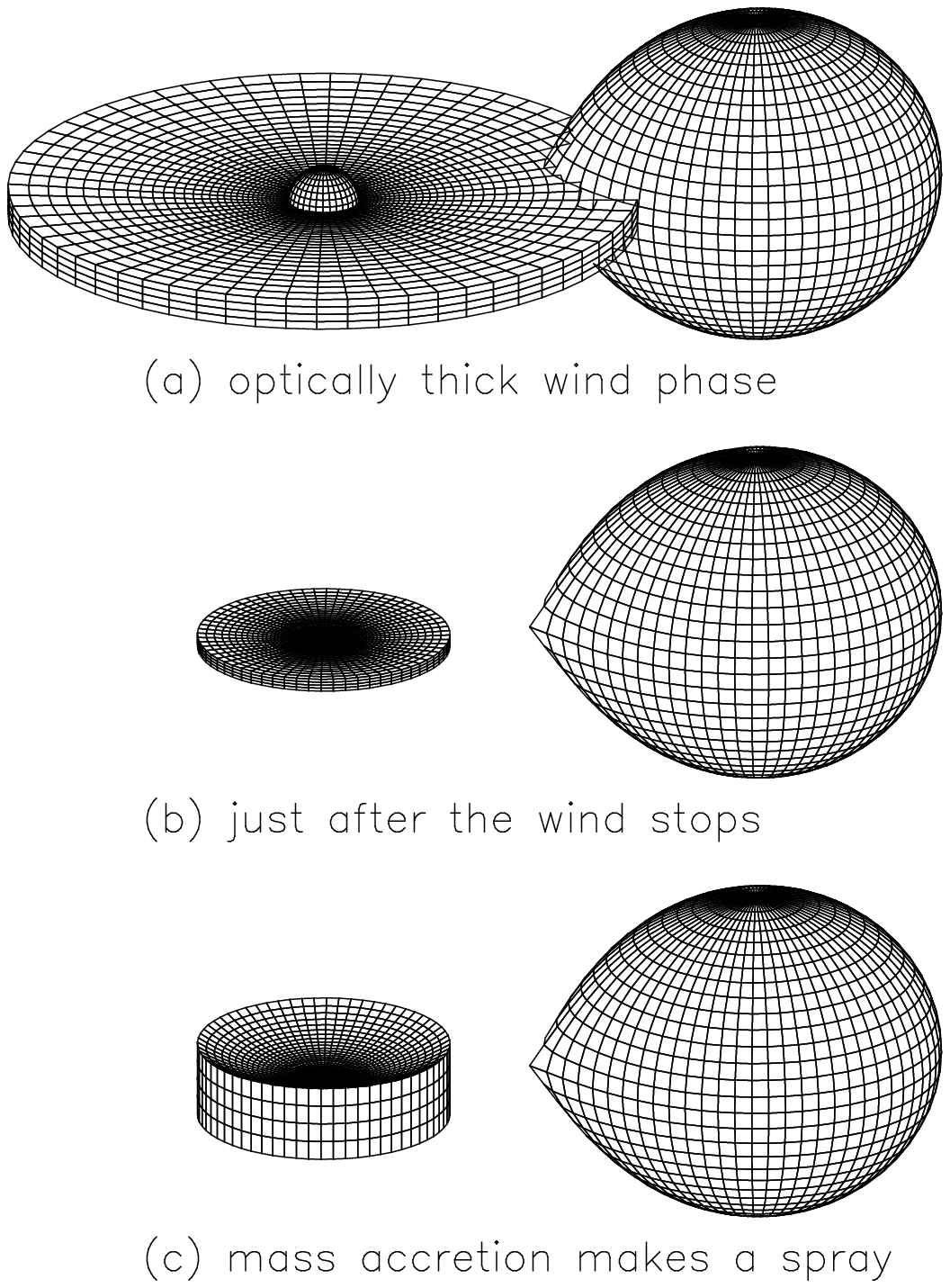}
\caption{
Configurations of our CI Aql model are illustrated: (a)
in the massive wind phase (X-ray off), (b) just after the wind stops
(X-ray on), and (c) during a rapid mass accretion phase 
soon after the wind stops (X-ray on).
The cool component ({\it right}) is a slightly evolved 
MS companion ($1.5 M_\odot$) filling its inner critical Roche lobe.  
The north and south polar areas of the cool component are 
irradiated by the hot component ($1.2~M_\odot$ white dwarf, {\it left}).
The separation is $a= 4.25 R_\odot$; 
the effective radii of the inner critical Roche lobes are
$R_1^*= 1.53 R_\odot$, and $R_2^*= R_2= 1.69 R_\odot$, 
for the primary white dwarf and the secondary main-sequence 
companion, respectively.
(a) In our model, we regard the outer edge of
the disk as the transition place from optically thick to thin.  
When the disk surface flow collides with the companion star,
it may run around the companion.  We approximate 
this situation simply by the disk configuration as shown 
in the figure.  This round flow hardly contributes
to light curves or line profiles because its irradiation effect
is much smaller than the other parts the disk. 
(b) The disk shrinks to a normal size ($0.7 - 0.8$ times the Roche lobe 
size) in several orbital periods after the wind stops.  
(c) A rapid mass accretion resumes and makes a spray around the disk edge.
\label{ci_aql_3figure}}
\end{figure}

\begin{figure}
\plotone{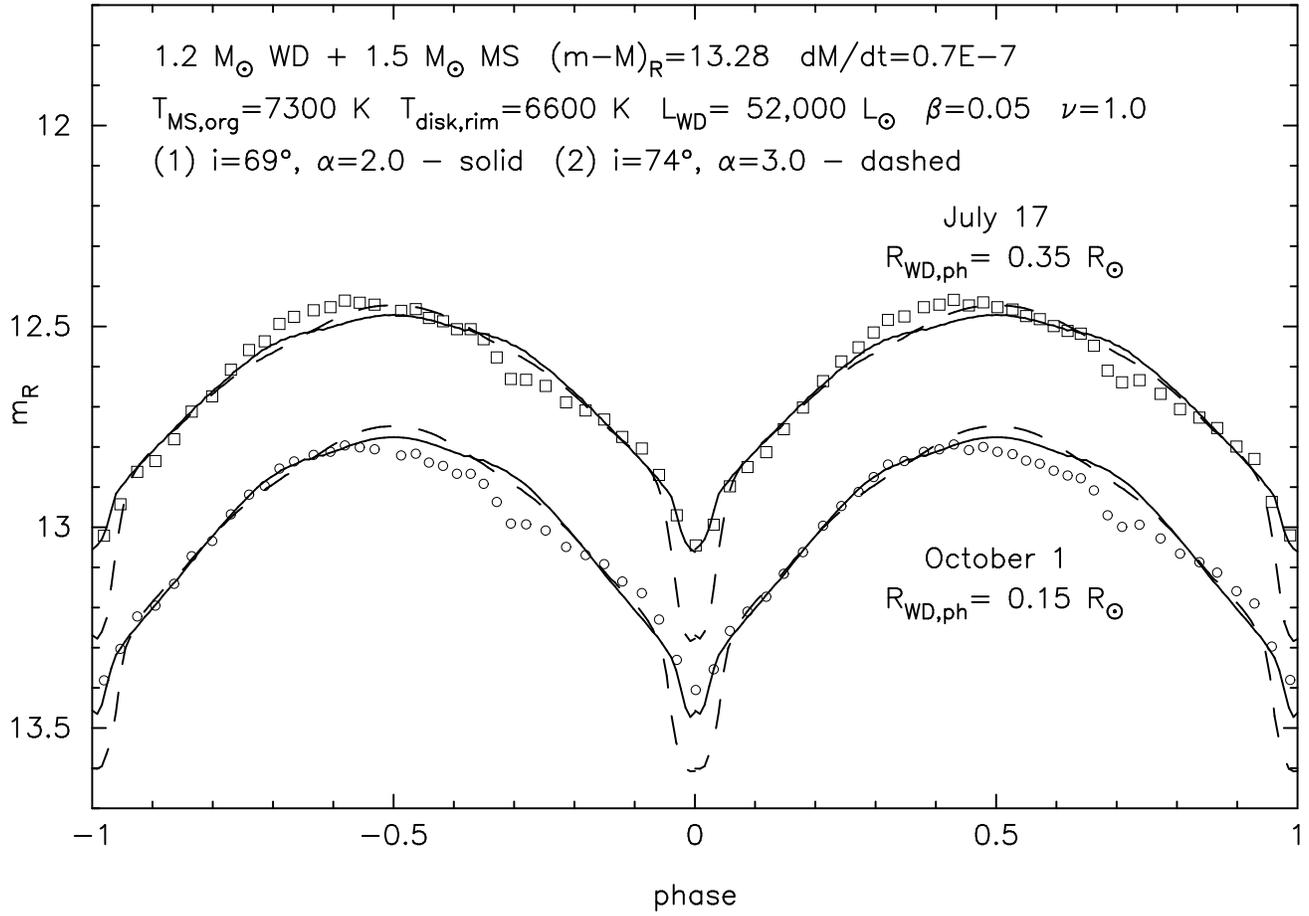}
\caption{
Calculated $R_c$ light curves are plotted against the binary phase 
(binary phase is repeated twice from $-1.0$ to $1.0$) together with
the observational points \citep{mat01} of the two epochs
(open squares correspond to the brightness on 2000 July 17
and open circles are the brightness on 2000 October 1).
Solid lines denote our calculated models with the inclination
angle of $i=69\arcdeg$ but dashed lines are of $i=74\arcdeg$.
Various system parameters are shown in the figure together with
the best fitted parameters.
\label{rc_mag_m15_wind_orbital}}
\end{figure}

     Here we plot two identical orbital modulations,
but one is shifted by 0.36 mag down.
The brighter one corresponds to the earlier stage
on 2000 July 17 while the fainter one does to that on 2000 October 1.
To reproduce the light curves with no secondary eclipses, we need both
(1) large (area) irradiation effect to reproduce the brightness, 
and (2) bright sources other than the companion to eliminate
the secondary eclipse.  Then, we assumed that (a) the outer edge of 
an optically thick disk extends up to almost the orbital separation
as illustrated in Figure \ref{ci_aql_3figure}a, 
and that (b) the WD photosphere is 
as large as $R_{\rm WD, ph} \sim 0.35~R_\sun$ on July 17 and
$R_{\rm WD, ph} \sim 0.15~R_\sun$ on October 1 if the other conditions
are the same between these two epochs.
Any light sources other than the disk or the companion
are unlikely to contribute to the $R_c$-magnitude because the strong
wind blows up off nebulosities around. 
This large extension of the disk naturally explains evident lack
of secondary minima.  The width of sharp eclipse minima,
$\sim 0.05 - 0.1$ orbital phase, is caused by the partial occultation 
of the WD photosphere.  
These large photospheric radii 
($R_{\rm WD, ph} \sim 0.35 \rightarrow 0.15~R_\sun$ 
from July 17 to October 1) are very consistent with our theoretical
model because WDs blow a wind when $R_{\rm WD, ph} > 0.085~R_\sun$
\citep[e.g.,][]{kat94h,hac01kb, hac02}.
\par
     The effect of a larger inclination
angle is canceled out by the effect of a larger radius of the disk.
For example, a combination of the inclination angle $i= 74\arcdeg$ 
and the disk of $(\alpha, \beta)=(3.0, 0.05)$ can also reproduce
the light curve modulations (dashed line) as shown in Figure 
\ref{rc_mag_m15_wind_orbital}.  
Its eclipse dip seems to be too deep to be compatible with 
the observation.  However, the total amplitude of the modulations is 
$\sim 0.8$ mag \citep[see Fig. 4 of][for the original observation]{mat01} 
and still consistent.  Here we adopted a set of $i= 74\arcdeg$
and $(\alpha, \beta)=(3.0, 0.05)$.
In both $i= 69\arcdeg$ and $i= 74\arcdeg$ cases, the secondary occults
only a part of the disk.  This is, however, enough to make a orbital
modulation observed because the central part of the disk is much brighter
than the outer part.
\par
     The velocities of winds calculated are as fast as $1000-1500$ 
km~s$^{-1}$ \citep[e.g.,][]{kat94h} and the wind mass loss rate is
as large as $10^{-6} - 10^{-4} M_\sun$~yr$^{-1}$ 
\citep[e.g.,][]{hac01kb}.  
Observationally much faster wind velocities are reported,
for example, $2000-2500$ km~s$^{-1}$ 
in CI Aql \citep[see, e.g.,][]{kis01}.
Such a strong wind affects the surface of the accretion disk.  
Because of a large velocity difference between the wind and the disk 
surface, it certainly drives the Kelvin-Helmholtz instability
at the interface.  Only the very surface layer of the disk is
dragged away outward with the velocity at least several hundred 
km~s$^{-1}$ like a free stream moving outward.
This surface free stream is optically thick
near the original disk but becomes optically thin somewhere
outside because of geometrical dilution effect.  We regard 
the transition place from optically thick to thin as the edge
of the extended disk.
\par
     It should be noted that a high density part of the disk is hardly
changed by this relatively tenuous surface free stream and 
still resides within its Roche lobe because the internal density
of the disk is much denser than that of the WD wind.
The wind mass loss rate is about $1 \times 10^{-6} M_\sun$~yr$^{-1}$
and its velocity is $\sim 1000$ km~s$^{-1}$, so that the density of 
the wind is estimated from the continuity ($\dot M_{\rm wind} = 
4 \pi r^2 \rho v$) to be about $10^{-11}$g~cm$^3$ at the distance 
of $1~R_\sun$ from the center of the WD.
On the other hand, the density of the standard accretion disk
is about $1 \times 10^{-3}$g~cm$^3$ at the same radius. 
Here, we assume the standard accretion disk model \citep{sha73}
with a mass accretion rate of $\sim 1 \times 10^{-7} M_\sun$~yr$^{-1}$.
\par
\begin{figure}
\plotone{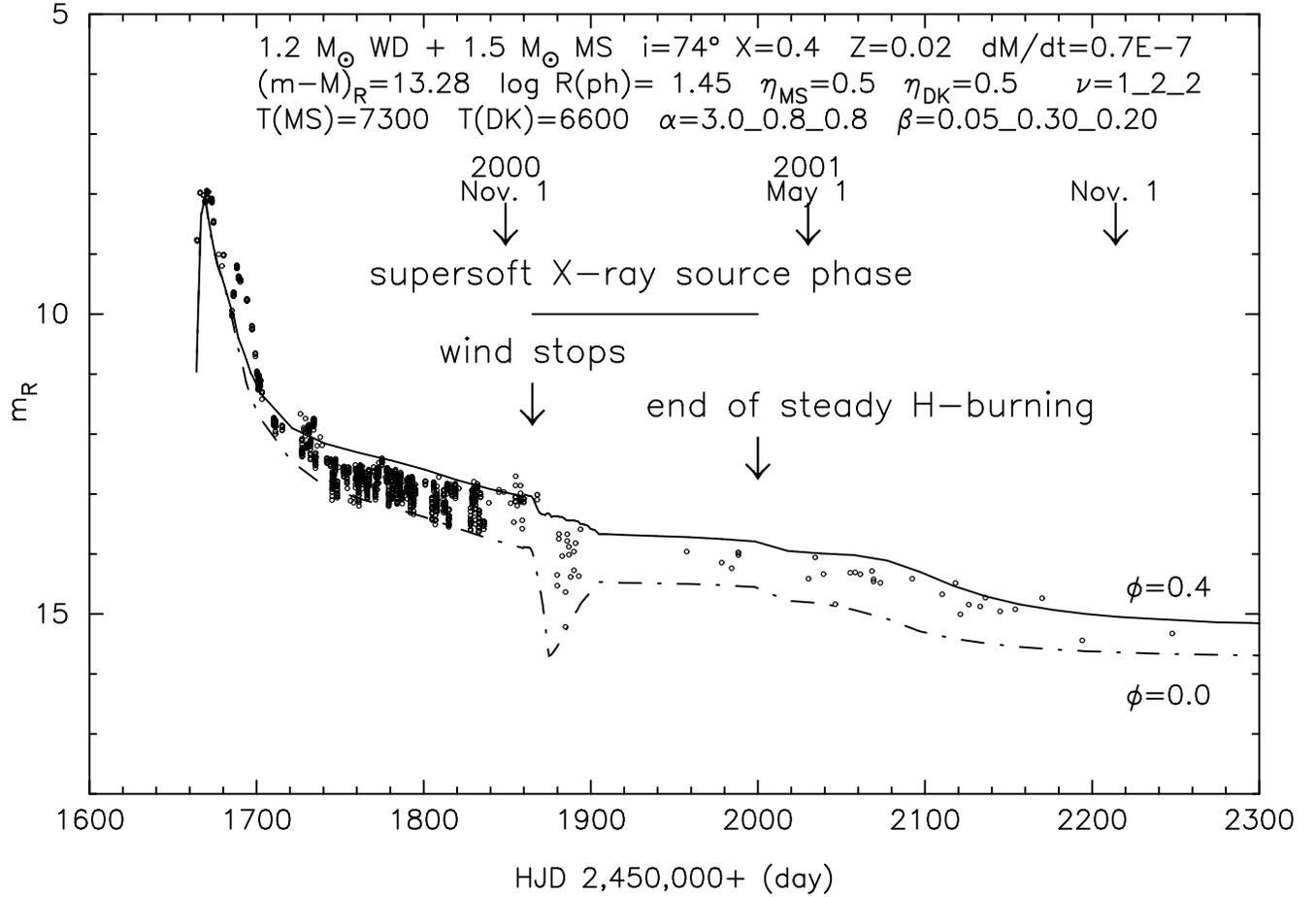}
\caption{
A calculated $R_c$-light curve are plotted against time (HJD 2,450,000+) 
together with the observational points of the CI Aql 2000 outburst
\citep[taken from][]{mat01, mat02}.
Small open circles indicate observational $R_c$-magnitudes.
The light curve connects the brightness at binary phase 0.4 (solid),
which is roughly corresponding to the upper bound of the light curve,
or at mid eclipse (binary phase 0.0) (dashed), which gives us 
the lower bound.  
The apparent distance modulus of $(m-M)_R= 13.28$ is used. 
The other system parameters are shown in the figure together with
the best fitted parameters. 
Here we adopt the hydrogen content of $X=0.4$ for refinement 
instead of $X=0.35$ by \citet{hac02}.
\label{rc_irradmix_ciaql00_m15}}
\end{figure}

 
     After the wind stops, the photospheric radius of the WD quickly 
shrinks in the Kelvin-Helmholtz timescale of the envelope, i.e.,
\begin{equation}
\tau_{\rm KH} \equiv {{\Delta E_{\rm env, thermal}} \over L}  
\sim {{2.3 \times 10^{44} {\rm erg}} \over
{2 \times 10^{38} {\rm erg~s}^{-1}}} \sim 13 {\rm ~days},
\label{KH_time_scale}
\end{equation}
where $\tau_{\rm KH}$ and $\Delta E_{\rm env, thermal}$ are
the Kelvin-Helmholtz timescale and the thermal energy of the WD 
envelope, respectively, and $L$ is the luminosity of the WD.
Here, we adopt a typical thermal energy of the WD envelope from
our solutions. 
It takes 20 days from $R_{\rm WD, ph} = 0.085~R_\sun$ to 
$0.04~R_\sun$ and takes another 20 days from $0.04~R_\sun$ 
to $0.02~R_\sun$ from our wind solutions and evolution models.
The disk edge of optically thick part 
shrinks to a normal size of $0.7 - 0.8$ times the Roche lobe radius
as observed in the supersoft X-ray sources \citep[e.g.,][]{hac02, led02}.
This may occur in several dynamical timescales, i.e., 
a few to several days.  Here, we adopt 10 days,
i.e., the transition from configuration (a) to (b) in Figure 
\ref{ci_aql_3figure}.  We adopt $(\alpha, \beta)=(0.8, 0.05)$ for 
configuration (b).  Then, the mass transfer rate from the MS to the WD
increases to make a spray around the accretion disk as seen in luminous
supersoft X-ray sources.  This can be mimicked 
by a flaring edge of the accretion disk, i.e., configuration (c) in 
Figure \ref{ci_aql_3figure}  \citep[see, e.g.,][for light curves and
flaring-up disk models after the wind stops]{hac02, led02}.  
We adopt $(\alpha, \beta)=(0.8, 0.3)$.  It also takes several dynamical
timescales.  Here we take $10 - 20$ days for the formation time of 
the spray.  Thus, we are able to reproduce a sharp $\sim 2$ mag drop 
around 2000 November 23 and the ensuing $\sim 1$ mag recovery soon 
after the drop, as shown in Figure \ref{rc_irradmix_ciaql00_m15}.  
A similar transition was observed in the U Sco 1999 outburst 
on HJD 2,451,244 just before the detection of supersoft X-rays 
\citep[see Fig. 2 of][]{hkkm00, kah99}.

\placefigure{rc_irradmix_ciaql00_m15}

\section{LMC supersoft X-ray source RX~J0513.9$-$6951}
     The LMC supersoft X-ray source RX~J0513.9$-$6951 shows regular
transitions between $60-120$ days optical high (X-ray off) 
and $30-40$ days optical low (X-ray on) state.
The X-ray off/optical high state was understood by 
a large expansion of the WD photosphere, which is
triggered by an increase in the mass transfer rate onto the
WD \citep[e.g.,][]{mey97, pak93, rei96, rei00, sou96}.  
When the WD envelope expands, an optically thick wind begins to blow.
As we already know in CI Aql, the sharp drop of $V$-mag is naturally
explained by cessation of the wind.  Our scenario is as follows:
\par\noindent
{\bf 1.} We assume the metallicity of $Z=0.004$ and hydrogen
content of $X=0.7$ for the envelope of the WD in RX~J0513 
since the LMC metallicity is reported to be about a third of
the solar metallicity. \\
{\bf 2.} We have directly calculated the duration of the wind phase and
the Kelvin-Helmholtz timescale on massive WDs of 1.0, 1.1, 1.2, 1.3,
1.35, 1.36, 1.37, and $1.377~M_\sun$ 
\citep[see also][for wind solutions]{hac01kb}.
The $1.3~M_\sun$ WD model 
gives us a reasonable wind timescale of $\tau_{\rm wind} \sim 120$ days
and a Kelvin-Helmholtz timescale of $\tau_{\rm KH} \sim 13$ days. \\
{\bf 3.}
We use the ephemeris revised by \citet{cow02}.
The inclination angle is not known but suggested to be
low.  Here we adopt $i= 10\arcdeg$.  The light curve is not so
different among $i=0\arcdeg - 20\arcdeg$.  The calculated light curve
is plotted in Figure \ref{vmagfit_rxj0513}, which is made 
by connecting brightness at the orbital phase 0.4 (phase 0.0 is
the minimum of the light curve modulation).  The mass ratio of 
$q=M_{\rm MS} / M_{\rm WD} =2$ is assumed because a high mass
transfer rate of $\gtrsim 10^{-6} M_\sun$~yr$^{-1}$ is suggested
\citep[e.g.][]{sou96}.  The non-irradiated surface temperature of
the lobe-filling 
MS companion is assumed to be 13,000 K \citep[e.g.,][]{fag94}. \\ 
{\bf 4.} We assume the original mass transfer rate of 
$2 \times 10^{-6} M_\sun$~yr$^{-1}$ as a reasonable value of 
thermally unstable mass transfer for $q=2$ and $M_{\rm MS}= 2.6~M_\sun$.
This value is realized whenever the wind does not suppress the mass
transfer.  So, we start a rapid mass accretion to 
the WD at the rate of $2 \times 10^{-6} M_\sun$~yr$^{-1}$ 
(denoted by $\square$-mark
in Fig. \ref{vmagfit_rxj0513}).  The WD envelope gradually 
expands and eventually blows a wind (see the mid and bottom panels of 
Fig. \ref{vmagfit_rxj0513}).  This causes a quick rise of $V$-mag.
Here we assume the disk expansion 
of $(\alpha, \beta )=(3.0,0.05)$ in the wind phase.  It takes a few to
several dynamical timescales.  We adopt 3 days.
The massive winds having the rate of 
$\dot M_{\rm wind} \sim 1 \times 10^{-8} - 1 \times 10^{-6} 
M_\sun$~yr$^{-1}$ certainly absorb supersoft X-rays 
\citep[e.g.,][]{sou96}.\\
{\bf 5.} This rapid mass accretion is suppressed by the strong wind,
because the wind hits and then strips-off the very surface layer of
the MS companion \citep[see][for a description of the stripping
effect by winds]{hkn99}. 
We stop the rapid mass accretion when the wind mass loss rate increases
and the mass-stripping rate overcomes the mass transfer rate of
$2 \times 10^{-6} M_\sun$~yr$^{-1}$.  This happens when the WD 
photosphere expands to the value shown in Figure \ref{vmagfit_rxj0513}
($\bigtriangleup$-mark).
However, we still keep a low rate 
($\sim 1 \times 10^{-7} M_\sun$~yr$^{-1}$)
during the period of draining of the disk 
\citep[a few times viscous timescale of the disk $\sim 40-60$ days, 
e.g.,][]{rei00} even after the rapid mass accretion stops. \\
{\bf 6.} When the optically thick winds stops ($\bigcirc$-mark), 
we expect copious supersoft X-rays as observed in the U Sco 1999
outburst \citep{kah99}.  Then the photosphere of the WD shrinks
from $0.087~R_\sun$ ($T_{\rm ph}= 280,000$~K) to $0.038~R_\sun$ 
(430,000~K) in $20$ days.  The photospheric temperature is 
consistent with the blackbody fitted temperatures ($30-40$ eV) 
of the supersoft X-rays \citep[e.g.,][]{sch93} as shown in Figure
\ref{vmagfit_rxj0513}.  The disk shrinks to a normal size 
of $(\alpha, \beta)=(0.8, 0.05)$ in a week and then the edge of
the disk begins to flare up, i.e., $(\alpha, \beta)=(0.8, 0.1)$ 
in another week, as described in \S 3. \\
{\bf 7.} Rapid mass accretion to the WD resumes ($\square$-mark)
$\sim 20$ days after the wind stops.  We expect that the mass transfer
rate gradually increases just after the wind stops but it takes 
$10-20$ days to reach the WD surface.
This timescale may be regarded as a viscous timescale of the accretion
disk \citep{rei00}.  Then, the photospheric radius of the white dwarf
expands in $20$ days and eventually blows winds. 
Then, go to {\bf 4} and repeat the cycle.
\par
Here we assume the apparent distance modulus of 
$(m-M)_V = 18.25$(intrinsic)$+ 0.4$(absorption)$=18.65$ to RX~J0513.
The calculated color of $(B-V)_c = -0.26$ is roughly consistent with 
the observational color of $(B-V)_o \sim -0.14 \pm 0.04$ \citep{pak93},
$(B-V)_o \sim -0.112 \pm 0.012$ \citep{cow93},
or $(B-V)_o \sim -0.04$ \citep{cow02} for the color 
excess of $E(B-V) \sim 0.13$ \citep{gan98}.

\placefigure{vmagfit_rxj0513} 

\section{DISCUSSION}
\begin{figure}
\plotone{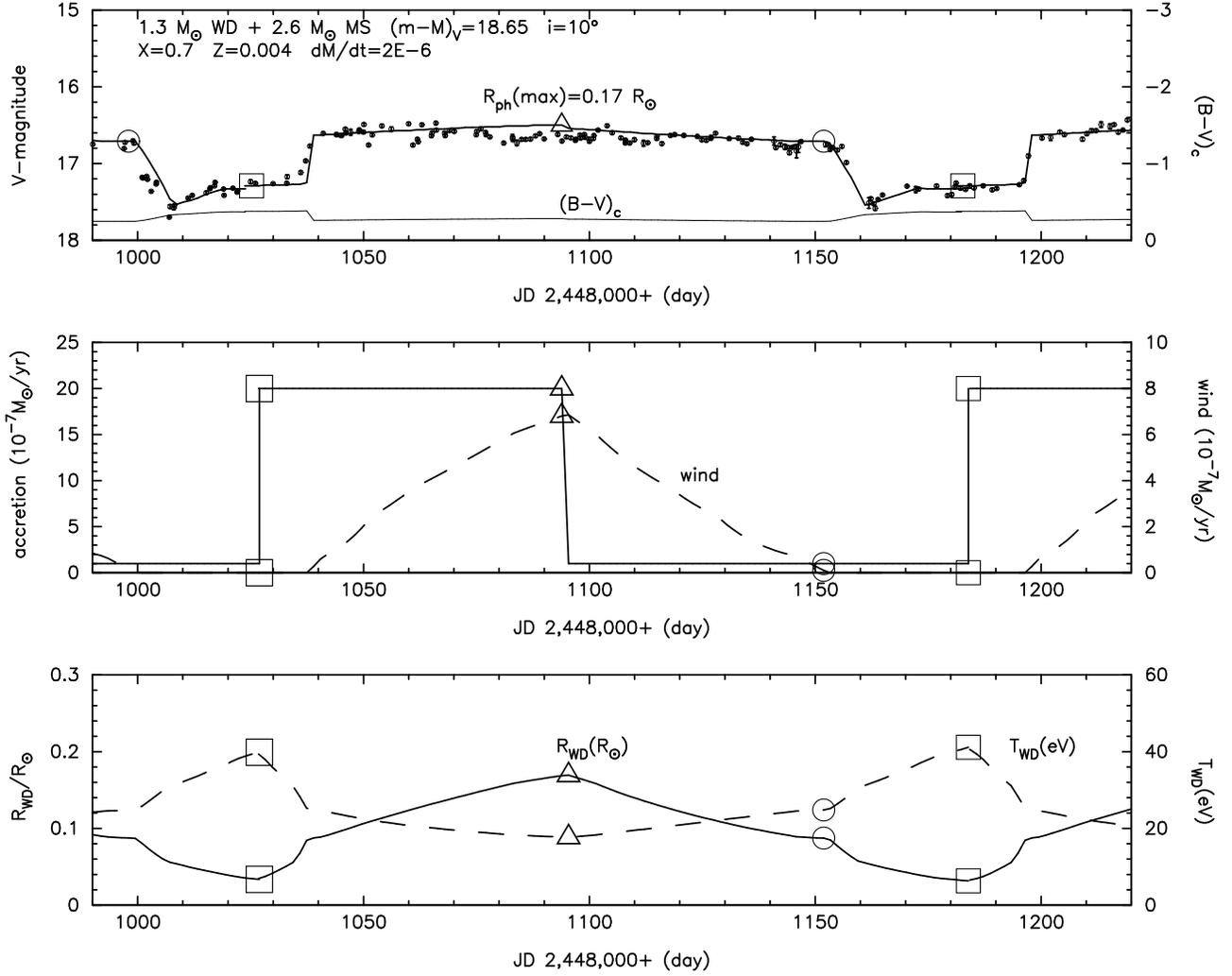}
\caption{
Top panel: $V$-magnitude light curves of RX~J0513.9$-$6951 
are plotted against time (JD 2,448,000$+$), together with 
the calculated color $(B-V)_c$ (lower portion).
Small open circles are observational data 
with an error bar \citep{alc96}.
Thick solid lines correspond to our calculated light curve, connecting
the brightness at orbital phase 0.4.  Thin solid lines denote the
calculated color $(B-V)_c$.
Starting a rapid mass accretion of $2 \times 10^{-6} M_\sun$~yr$^{-1}$
to the WD at epoch of open square, we continue it until epoch of open
triangle.  Optically thick winds stop at epoch of large open circle.
The maximum expansion of the WD is reached at epoch of open
triangle, the radius of which is shown above the open triangle.
Mid panel: The accretion rate to the WD (solid line) 
and the wind mass loss rate from the WD (dashed line) 
are plotted against time.  The other symbols have the same meaning
as those in the top panel.  
Third panel: The white dwarf radius (solid line) and photospheric
temperature (dashed line) are plotted against time.  The other
symbols are the same as those in the top panel.
\label{vmagfit_rxj0513}}
\end{figure}

\begin{figure}
\plotone{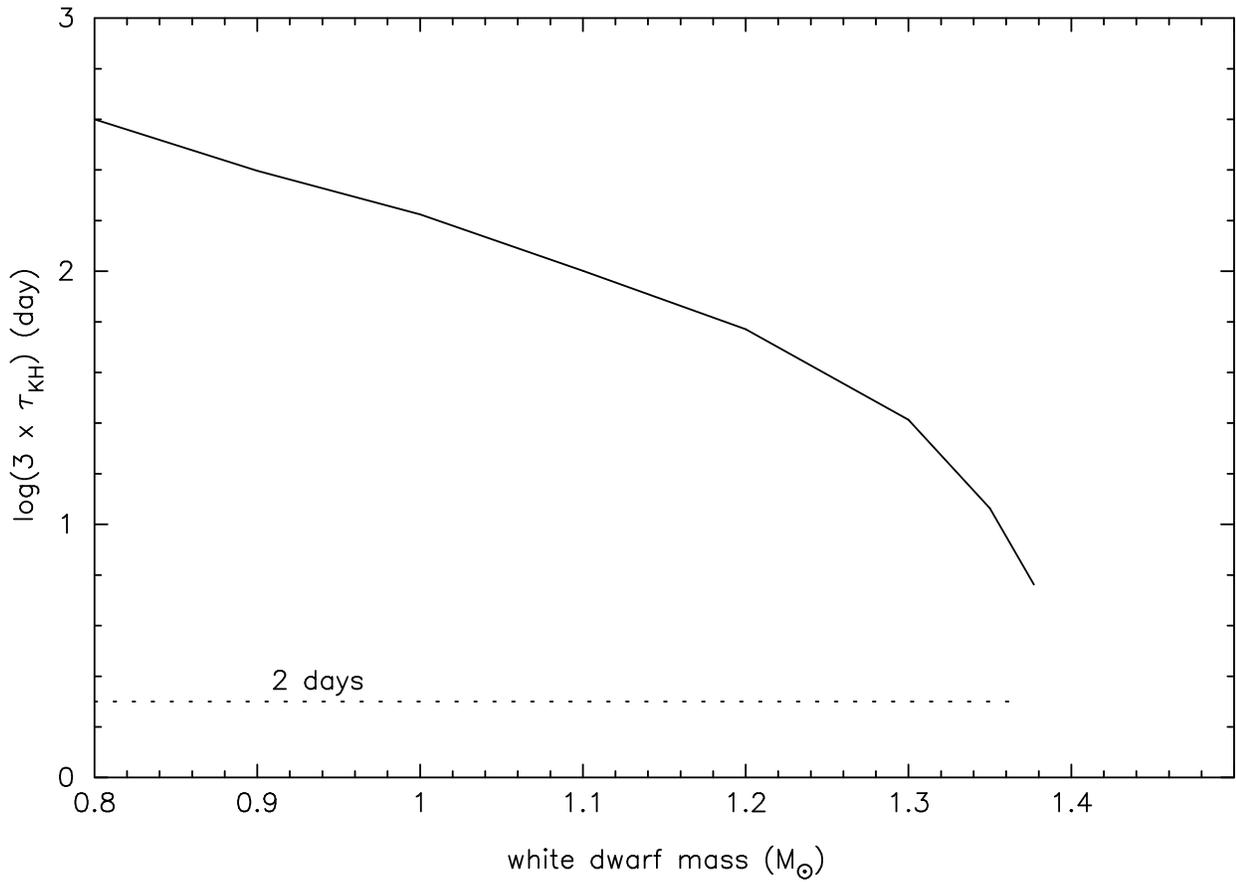}
\caption{
Timescale of contraction/expansion of white dwarf envelope 
with $Z=0.004$ and $X=0.7$ is plotted against the white dwarf mass.
Here, we adopt time during which the white dwarf photosphere shrinks
from $0.08~R_\sun$ to $0.02~R_\sun$ by a factor of four,
which is about three times the Kelvin-Helmholtz timescale of
the white dwarf envelope. 
\label{thermal_time_one}}
\end{figure}

     Thus, we are able to reproduce the transition between 
the optical high and low states of RX~J0513 as shown in Figure 
\ref{vmagfit_rxj0513}.  Our new model is essentially different from
the previous expansion/contraction models of the white dwarf
photosphere in the following sense:  The emergence/decay timescale of
the observed supersoft X-rays is as short as one or two days
\citep{rei00}.  This short timescale cannot be explained by
the expansion/contraction of the envelope.
We have calculated the timescales of photospheric
contraction based on our WD envelope model \citep{kat94h} and
show them in Figure \ref{thermal_time_one} against the WD mass.
The contraction timescale depends on the envelope mass of the WD and
it is shortest just after the optically thick wind stops as shown
in Figure \ref{vmagfit_rxj0513}.  In this phase, the timescale of 
contraction by a factor of four is about three times 
the Kelvin-Helmholtz timescale.  It is clear that the timescale of
contraction by a factor of four cannot be shorter than six days
even if the WD mass is very close to the Chandrasekhar mass.
We may conclude that only the expansion/contraction
model of the white dwarf photosphere cannot reproduce 
a very short timescale of X-ray on/off.  We need another mechanism. 
\par
     On the contrary, optically thick winds completely obscure 
supersoft X-rays within one day after it starts.  The wind mass loss
rate increases from zero to $\sim 2 \times 10^{-8} M_\sun$~yr$^{-1}$
in a day when it starts, or it decreases from 
$\sim 1 \times 10^{-8} M_\sun$~yr$^{-1}$ to zero in a day when it ceases,
as easily seen from Figure \ref{vmagfit_rxj0513}.
This wind mass loss rate of $1 \times 10^{-8} M_\sun$~yr$^{-1}$
is large enough to absorb supersoft X-rays
\citep[see, e.g., discussion of][]{sou96}.
\par
     In the scenario proposed by \citet{hkn99, hknu99}, 
all the progenitor systems of Type Ia supernovae once experience
the optically thick wind phase during their way to Type Ia supernova.
RX~J0513 may be an example for such a wind phase. 
We have calculated the efficiency of mass accretion to be 
about 35\% for the case of Figure \ref{vmagfit_rxj0513}. 
The other 65\% is blown in the wind or stripped off from 
the companion surface and eventually lost from the binary system.
Assuming this efficiency being kept, we can expect
that the white dwarf in RX~J0513.9$-$6951 will explode 
as a Type Ia supernova after about $0.6 M_\sun$ is transferred
even when the WD mass is $1.2~M_\sun$.

\placefigure{thermal_time_one}



\acknowledgments
     We thank K. Matsumoto for providing us machine readable
data sets of the CI Aql 2000 outburst.
We are also greatly indebted to F. Meyer and Emi Meyer-Hofmeister
for stimulating us to work on RX~J0513.9$-$6951 during our stay
at the Max-Planck Institute for Astrophysics.
This research has been supported in part by the Grant-in-Aid for
Scientific Research (11640226) 
of the Japan Society for the Promotion of Science.

\clearpage

\clearpage

\clearpage

\clearpage

\clearpage



\begin{thebibliography}{}
\bibitem[Alcock et al. (1996)]{alc96}
Alcock, C. et al. 1996, \mnras, 280, L49



\bibitem[Cowley et al. (2002)]{cow02}
Cowley, A.P., Schmidtke, P.C., Crampton, D., \& Hutchings, J.B. 2002, 
\aj, 124, 2233

\bibitem[Cowley et al. (1993)]{cow93}
Cowley, A. P., Schmidtke, P. C., Hutchings, J. B., Crampton, D., \& 
McGrath, T. K. 1993, \apj, 418, L63

\bibitem[Fagotto et al. (1994)]{fag94}
Fagotto, F., Bressan A., Bertelli G., Chiosi C. 1994, \aaps, 105, 29

\bibitem[G\"ansicke et al. (1998)]{gan98}
G\"ansicke, B. T., van Teeseling, A., Beuermann, K., \& de Martino, D.
1998, \aap, 333, 163


\bibitem[Greiner \& DiStefano (2002)]{gre02}
Greiner, J., \& DiStefano, R. 2002, \apj, in press (astro-ph/0210035)

\bibitem[Hachisu \& Kato (2001a)]{hac01ka}
Hachisu, I., \& Kato, M. 2001a, \apjl, 553, L161

\bibitem[Hachisu \& Kato (2001b)]{hac01kb}
Hachisu, I., \& Kato, M. 2001b, \apj, 558, 323

\bibitem[Hachisu et al. (2000)]{hkkm00}
Hachisu, I., Kato, M., Kato, T., \& Matsumoto, K. 2000, 
\apjl, 528, L97 



\bibitem[Hachisu et al. (1999a)Hachisu, Kato, \& Nomoto]{hkn99}
Hachisu, I., Kato, M., \& Nomoto, K. 1999a, \apj, 522, 487 

\bibitem[Hachisu et al. (1999b)]{hknu99}
Hachisu, I., Kato, M., Nomoto, K., \& Umeda, H. 1999b, \apj, 519, 314 

\bibitem[Hachisu et al. (2002)Hachisu, Kato, \& Schaefer]{hac02}
Hachisu, I., Kato, M., \& Schaefer, B. E. 2002, \apj, in press
(astro-ph/0210592)
%


\bibitem[Kahabka et al. (1999)]{kah99}
Kahabka, P., Hartmann, H. W., Parmar, A. N., \& Negueruela, I. 1999, 
\aap, 374, L43


\bibitem[Kato \& Hachisu (1994)]{kat94h}
Kato, M., \& Hachisu, I., 1994, \apj, 437, 802


\bibitem[Kiss et al. (2001)]{kis01}
Kiss, L. L., Thomson, J. R., Ogloza, W., F\"ur\'esz, G., \& 
Szil\'adi, K. 2001, \aap, 366, 858

\bibitem[Lederle \& Kimeswenger (2002)]{led02}
Lederle, C., \& Kimeswenger, S. 2002, \aap, in press (astro-ph/0209580)



\bibitem[Matsumoto et al. (2002)]{mat02}
Matsumoto, K., Ishioka, R., Uemura, M., Kato, T., Kawabata, T. 2002, 
\mnras, in press (astro-ph/0211241)

\bibitem[Matsumoto et al. (2001)]{mat01}
Matsumoto, K., et al. 2001, \aap, 378, 487


\bibitem[Meyer-Hofmeister et al. (1997)Meyer-Hofmeister, Schandl, 
\& Meyer]{mey97}
Meyer-Hofmeister, E., Schandl, S., \& Meyer, F. 1997, \aap, 321, 245


\bibitem[Nomoto et al. (1984)Nomoto,  Thielemann, \& Yokoi]{nom84}
Nomoto,  K., Thielemann, F., \& Yokoi, K. 1984, \apj, 286, 644

\bibitem[Pakull et al. (1993)]{pak93}
Pakull, M. W., Motch, C., Bianchi, L., Thomas, H.-C., Guibert, J., 
Beaulieu, J. P., Grison, P., \& Schaeidt, S. 1993, \aap, 278, L39




\bibitem[Reinsch et al. (1996)]{rei96}
Reinsch, K., van Teeseling, A., Beuermann, K., \& Abbott, T. M. C. 1996,
\aap, 309, L11

\bibitem[Reinsch et al. (2000)]{rei00}
Reinsch, K., van Teeseling, A., King, A. R., \& Beuermann, K. 2000,
\aap, 354, L37




\bibitem[Schaeidt et al. (1993)Schaeidt, Hasinger, \& Truemper]{sch93}
Schaeidt, S., Hasinger, G., \& Truemper, J. 1993, \aap, 270, L9

\bibitem[Schandl et al. (1997)Schandl, Meyer-Hofmeister, \& Meyer]{sch97}
Schandl, S., Meyer-Hofmeister, E., \& Meyer, F. 1997, \aap, 318, 73

\bibitem[Schmeja et al. (2000)Schmeja, Armsdorfer, \& Kimeswenger]{sch00}
Schmeja, S., Armsdorfer, B., \& Kimeswenger, S. 2000, 
Inf. Bull. Variable Stars, 4957

\bibitem[Shakura \& Sunyaev (1973)]{sha73}
Shakura, N. I., \& Sunyaev, R. A. 1973, \aap, 24, 337

\bibitem[Southwell et al. (1996)]{sou96}
Southwell, K. A., Livio, M., Charles, P. A., O'Donoghue, D., 
\& Sutherland, W. J. 1996, \apj, 470, 1065




\bibitem[Thoroughgood et al. (2001)]{tho01}
Thoroughgood, T. D., Dhillon, V. S., Littlefair, S. P., Marsh, T. R.,
\& Smith, D. A. 2001, \mnras, 327, 1323



%
\end{thebibliography}
\end{document}